\documentclass[12pt]{article}

\usepackage{scicite} 
\usepackage{graphicx} 
\usepackage{times}   
\usepackage{enumitem}
\usepackage{makecell} 
\usepackage{url}

\def\aap{{Astron. Astrophys.}}

\def\apjl{{Astrophys. J. Lett.}}
\def\apjs{{Astrophys. J. Suppl. Ser.}}

\newcommand{\Nflares}{2889 }
\newcommand{\Nflaringstars}{2527 }
\newcommand{\NstarinputBeforeFastRotatorsCleaning}{56,496 }
\newcommand{\Nstarinputfinal}{56,450 }
\newcommand{\NstarinputfinalNonPeriodic}{39,347 }
\newcommand{\NstarinputfinalPeriodic}{17,103 }
\newcommand{\nfast}{46 }
\newcommand{\nfastper}{37 }
\newcommand{\NumberOfFlaresInDaysWithTooLargeFlareRate}{373 }
\newcommand{\NumberOfFLaresDetectedWithOkamotoThreshold}{1376 }

\newcommand{\Teff}{T_{\rm{eff}}}

\newcommand{\Prot}{P_{\rm{rot}}}
\newcommand{\Rvar}{R_{\rm var}}

\newenvironment{sciabstract}{%
\begin{quote} \bf}
{\end{quote}}

\title{Sun-like stars produce superflares roughly once per century} 

\author{Valeriy Vasilyev,$^{1\ast}$ 
Timo Reinhold,$^{1}$ 
Alexander I.\ Shapiro,$^{1,2}$
Ilya Usoskin,$^{3, 4}$ \\
Natalie A.\ Krivova,$^{1}$ 
Hiroyuki Maehara,$^{5}$
Yuta Notsu,$^{6,7}$ 
Allan Sacha Brun,$^{8}$ \\
Sami K.\ Solanki,$^{1}$
Laurent Gizon$^{1,9}$
\\
\normalsize{
$^{1}$Max-Planck-Institut f\"ur Sonnensystemforschung,} \\
\normalsize{37077 G\"ottingen, Germany}\\
\normalsize{$^{2}$Institute of Physics, University of Graz,  8010 Graz, Austria}\\
\normalsize{
$^{3}$Sodankyl\"a Geophysical Observatory, University of Oulu, 90014 Oulu, Finland}\\
\normalsize{$^{4}$ Space Physics and Astronomy Research unit, University of Oulu, 90014 Oulu, Finland}\\
\normalsize{
$^{5}$Subaru Telescope Okayama Branch Office, National Astronomical Observatory of Japan,}\\
\normalsize{Okayama 719-0232, Japan}\\
\normalsize{
$^{6}$Laboratory for Atmospheric and Space Physics, University of Colorado Boulder,}\\
\normalsize{Boulder, CO 80303, USA}\\
\normalsize{
$^{7}$National Solar Observatory, Boulder, CO 80303, USA}\\
\normalsize{
$^{8}$  Laboratoire Astrophysique Instrumentation and Modélisation, }\\
\normalsize{Universities of Paris-Cité and Paris-Saclay, Center for Atomic and Alternative Energies,}\\ 
\normalsize{Center of National Research in Science, 91191 Gif-sur-Yvette, France}\\
\normalsize{$^{9}$Institut für Astrophysik, Georg-August-Universit\"at G\"ottingen,  37077  G\"ottingen, Germany}\\
\normalsize{$^\ast$Corresponding author. E-mail:  vasilyev@mps.mpg.de.}
}

\date{}

\begin{document} 


\baselineskip24pt


\maketitle


\begin{sciabstract}
Stellar superflares are energetic outbursts of electromagnetic radiation, similar to solar flares but releasing more energy, up to $10^{36}$~erg on main sequence stars. 
It is unknown whether the Sun can generate superflares, and if so, how often they might occur. We used photometry from the Kepler space observatory to investigate superflares on other stars with Sun-like fundamental parameters. We identified  \Nflares superflares on \Nflaringstars Sun-like stars, out of 56,450 observed. This detection rate indicates that superflares with energies $>10^{34}$~erg occur roughly once per century on stars with Sun-like temperature and variability. The resulting stellar superflare frequency-energy distribution is consistent with an extrapolation of the Sun’s flare distribution to higher energies, so we suggest that both are generated by the same physical mechanism.
\end{sciabstract}

Solar flares are sudden local bursts of bright electromagnetic emission from the Sun, which release a large amount of energy within a short interval of time \cite{Benz2017}.
The increase in short-wavelength solar radiation during flares influences the Earth's upper atmosphere and ionosphere, sometimes causing radio blackouts and ionosphere density changes \cite{Mitra1974}.
Solar flares are frequently accompanied by the expulsion of large volumes of plasma, known as coronal mass ejections (CMEs), which accelerate charged particles to high energies. When these solar energetic particles (SEPs) reach Earth, they cause radiation hazards to spacecraft, aircraft and humans. 
Extreme SEP events can produce isotopes, called cosmogenic isotopes, which form when high-energy particles interact with the Earth’s atmosphere. These isotopes are then recorded in natural archives, such as tree rings and ice cores \cite{miyake19,usoskin_LR_23}.
The total amount of energy released by each flare varies by many orders of magnitude, as determined by a complex interplay between the physical mechanisms of particle acceleration and plasma heating in the Sun's atmosphere \cite{gopalswamy18}.

Solar flares have been observed for less than two centuries. 
Although thousands of them have been detected and measured \cite{Veronig2002, Aschwanden2012ApJ, Plutino2023}, only about a dozen are known to have exceeded a bolometric  (integrated over all wavelengths) energy of $10^{32}$~erg \cite{Emslie2012ApJ, Cliver2022}.
Among them was the Carrington Event on 1 September 1859 \cite{Carrington1859,Hodgson1859}, which was accompanied by a CME that  had the strongest recorded impact on Earth. Modern estimates of the Carrington Event's total bolometric  energy are $4\times 10^{32}$ to $6 \times  10^{32}$~erg \cite{Cliver2013,Hayakawa2023}.

It is unknown whether the Sun can unleash flares with even higher energies, often referred to as superflares,
and if so, how frequently that could happen. The period of direct solar observations is too short to reach any firm conclusions. There are two indirect methods to investigate the potential for more intense flares on the Sun. One method uses extreme SEP events recorded in cosmogenic isotope data \cite{miyake12,Cliver2022}, which have been used to quantify the occurrence rate of strong CMEs reaching Earth over the past few millennia \cite{usoskin_LR_23}. 
There are five confirmed (and three candidate) extreme SEP events that are known to have occurred in the last 10,000 yr  \cite{usoskin_LR_23}, 
implying a mean occurrence rate of $\sim10^{-3}$ yr$^{-1}$. However, the relationship between SEPs and flares is poorly understood, especially for the stronger events \cite{Cliver2022}. 

A second method is to study superflares on stars similar to the Sun.
If the properties of the observed stars sufficiently match the Sun, the superflare occurrence rate on those stars can be used to estimate the rate on the Sun.  Studies using this method have indicated that superflares with energies of about $10^{34}$~erg occur with a frequency of $\sim (1.25 \pm 0.87) \times 10^{-3}$~yr$^{-1}$ on Sun-like stars \cite{Maehara2012}; all uncertainties are $1\sigma$. However, other studies have found substantially lower occurrence rates, probably because of differences in the selection of stars with Sun-like rotation rates \cite{Notsu2019,Okamoto2021}. For example, one study \cite{Okamoto2021} found 26 flares on 15 Sun-like stars (out of 1641 observed), implying an occurrence rate of 
$\sim (3.33\pm 1.25) \times 10^{-4}$  yr$^{-1}$ for a $7 \times 10^{33}$~erg superflare.

Those previous studies were limited to stars with known rotational periods \cite{McQuillan2014}. However, the majority of stars with measurable rotation periods are much more variable than the Sun \cite{Reinhold2020}, potentially biasing the results. Detecting the rotation period of a Sun-like star is challenging using standard methods \cite{Eliana2020, Reinhold2021}.
Therefore, the stars that are most similar to the Sun would have unknown rotation periods and thus were excluded from previous flare studies.


\paragraph*{Kepler observations of superflares.} We searched for superflares on Sun-like stars observed by the Kepler space telescope. 
We chose to include stars with unknown rotation periods in the analysis \cite{suppl}. Our sample consists of main-sequence stars with near-solar fundamental parameters, selected on the basis of effective temperatures $5000$\,K $<\Teff<$ $6500$\,K and G-band absolute magnitudes $4$\, mag $<M_G<$ $6$\,mag from the Kepler archive \cite{ KeplerDr25}, supplemented with data from Gaia Data Release 3 (DR3) \cite{GaiaDR3}. For context, the Sun has an effective temperature  $T_{\mathrm{eff}}^{\odot}=5780$~K and an absolute magnitude  $M_G^{\odot}=4.66$ mag.
We used Gaia astrometry to exclude unresolved binary systems with semi-major axes greater than 0.1 astronomical units \cite{Belokurov2020, suppl}. 
We excluded stars with known rotation periods shorter than 20 days \cite{suppl} using a rotation period catalog \cite{Reinhold2023}.  These stars are likely younger than the Sun (which has a rotation period of 25 days) and therefore more active, resulting in a higher flare frequency, which could bias the measured superflare frequency.

\begin{figure*}
    \centering
    \includegraphics[width=1.0\textwidth]{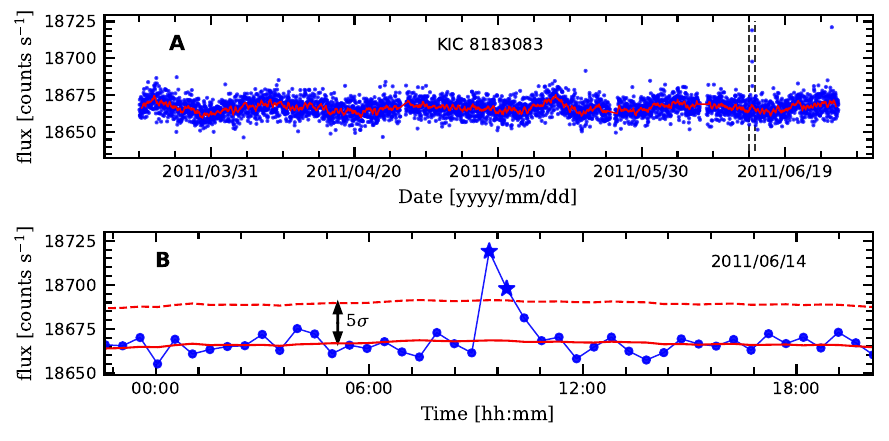}
    \caption{ {\bf Example superflare light curve.} \textbf{(A)} Observed light curve of the Sun-like star KIC\,8183083 (blue dots) and the same data smoothed with a 7.5 hr window (red curve). The vertical dashed lines enclose a superflare identified by our automated algorithm.  \textbf{(B)} Zoom into the region between the dashed lines in panel A. Star symbols indicate data points $>5\sigma$ above (dashed red line) the smoothed light curve. The uncertainties in the flux values are smaller than the size of the data points.}
    \label{fig:Figure1}
\end{figure*}

\begin{figure*}
    \centering
    \includegraphics[width=1.0\textwidth]{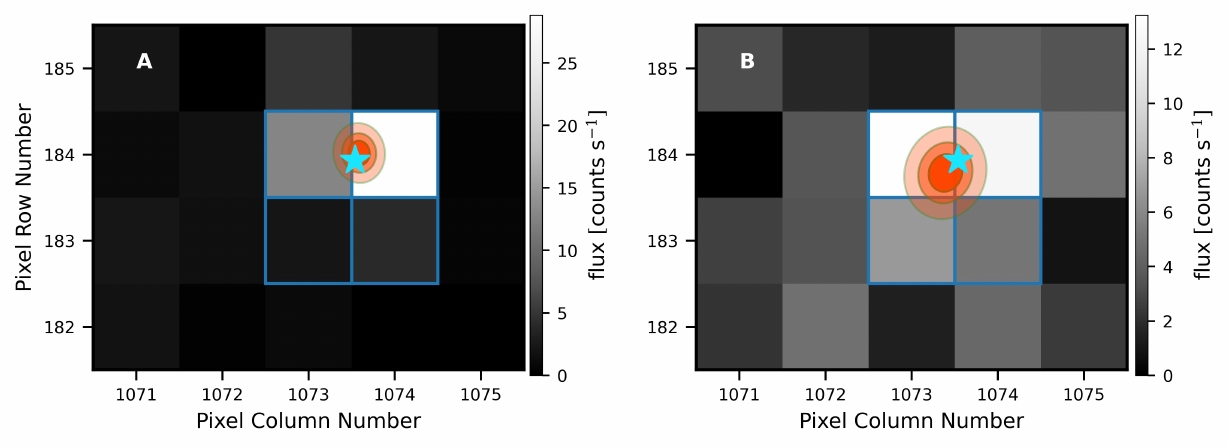}
    \caption{ {\bf Spatial localization of the example superflare.} 
    \textbf{(A)} Kepler image of the same flare on KIC 8183083 as in Fig.~1 during the maximum flare light.  \textbf{(B)} The same flare 30 min later. Greyscale (see color bar) indicates the flux in each pixel during the flare. The blue outlines indicate the four pixels used to extract the light curve (Fig. 1) and the cyan star is the catalog position of the star \cite{GaiaDR3}. The orange ellipses indicate the 68\%, 95\%, and 99.9\% confidence levels obtained by fitting a model of a point source to those images. The star is located within the 99.9\% confidence contour in both images, so we attribute the flare in the light curve to this target.}
\label{fig:Figure2}
\end{figure*}

We applied an automated flare detection algorithm \cite{Vasilyev2022} to these stars. The method uses a combined analysis of the light curve (observed stellar brightness as a function of time) and images to identify flare locations on the detector with subpixel precision. 
All stars in our sample are observed by the Kepler mission with a 30-min exposure time, so a typical superflare profile consisted only of a few data points (Fig.~\ref{fig:Figure1}). To detect superflares with high statistical confidence, we searched for events with at least two consecutive data points that exceed the background by at least a  $5\sigma$ threshold.  Each data point in the stellar light curve corresponds to the photon flux integrated over a particular pixel mask around the target star. In some cases, the light curve is affected by events not related to the target star, such as cosmic rays, flares on (un)resolved background stars, and minor Solar System bodies passing through the field of view.  Although we expect such contamination to be rare, it might strongly affect the flare statistics of stars with low flare occurrence rates.  Therefore, we additionally analyzed the images \cite{suppl} corresponding to the first two data points exceeding the 5$\sigma$ threshold in each light curve to spatially localize the flare source (Fig.~\ref{fig:Figure2}). Our algorithm fits a model of a point source to those images  and localizes the flare on a subpixel level. If the target Sun-like star resides within the 99.9\% confidence ellipse of the probable flare location, we attribute the flare to that star.

We examined the light curves of stars with multiple flares to exclude potential contamination of our sample by fast rotators \cite{suppl} that were not reported in the catalog  \cite{Reinhold2023}.
In total, we analyzed \Nstarinputfinal stars, of which \NstarinputfinalNonPeriodic stars had unknown rotation periods and \NstarinputfinalPeriodic had measured rotation periods \cite{Reinhold2023}; from that sample we identified \Nflares flares on \Nflaringstars stars. The bolometric flare energies range from $\sim10^{33}$ to $\sim10^{36}$~erg \cite{suppl}. With four years of observations for each star, our full sample corresponds to $\approx 220,000$ years of stellar activity - roughly 18 times longer than the cosmogenic isotope record of the Sun [$\approx 12,000$ years \cite{usoskin_LR_23}].

\begin{figure*}
\centering
    \includegraphics[width=1.0\textwidth]{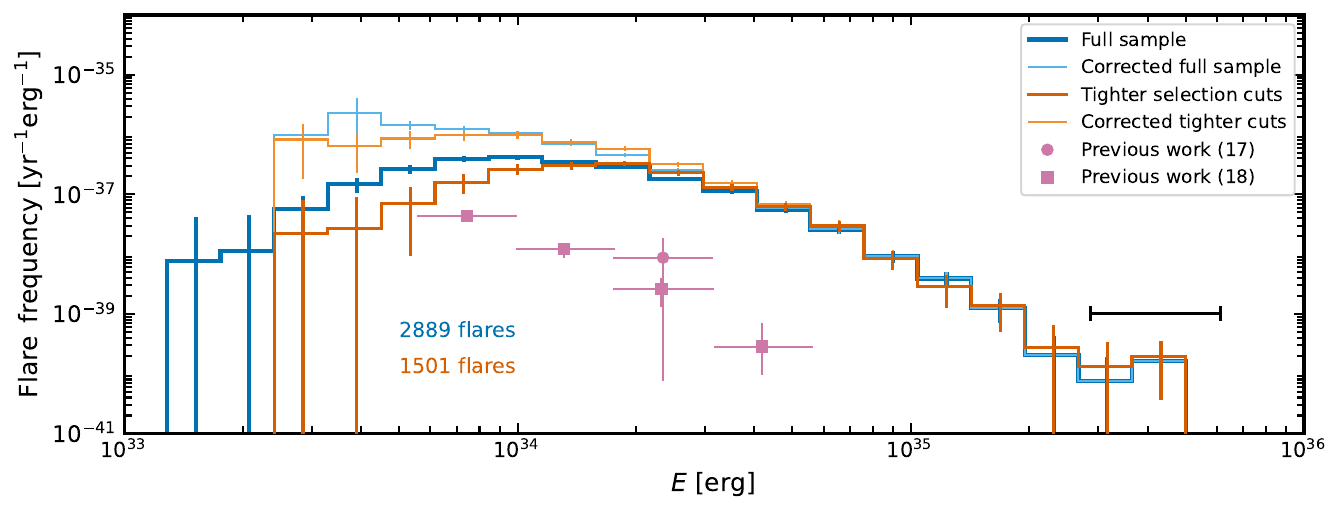}
    \caption{ \textbf{Frequency-energy distribution of stellar flares.} The thick blue histogram shows the flare frequency - the number of flares per star per year per unit energy - of all stars in our full sample  as a function of the flare energy. The thick orange  histogram shows the results for our sample with tighter selection cuts (effective temperatures 5500 to 6000\,K and variabilities $\Rvar < 0.3\%$).
    The thin blue and thin orange curves show the same distributions after correction for the missing low-energy flares due to the detection threshold \cite{suppl}. The error bars on each histogram bin show $1\sigma$ uncertainties \cite{suppl}. 
    Results from previous studies are shown as pink  circles \cite{Notsu2019} and  squares \cite{Okamoto2021}. The black horizontal errorbar indicates the typical $1\sigma$ uncertainty in stellar flare energy.}
\label{fig:pdf_flare_occurence_energy}
\end{figure*}

\begin{figure*}
\centering
    \includegraphics[width=1.0\textwidth]{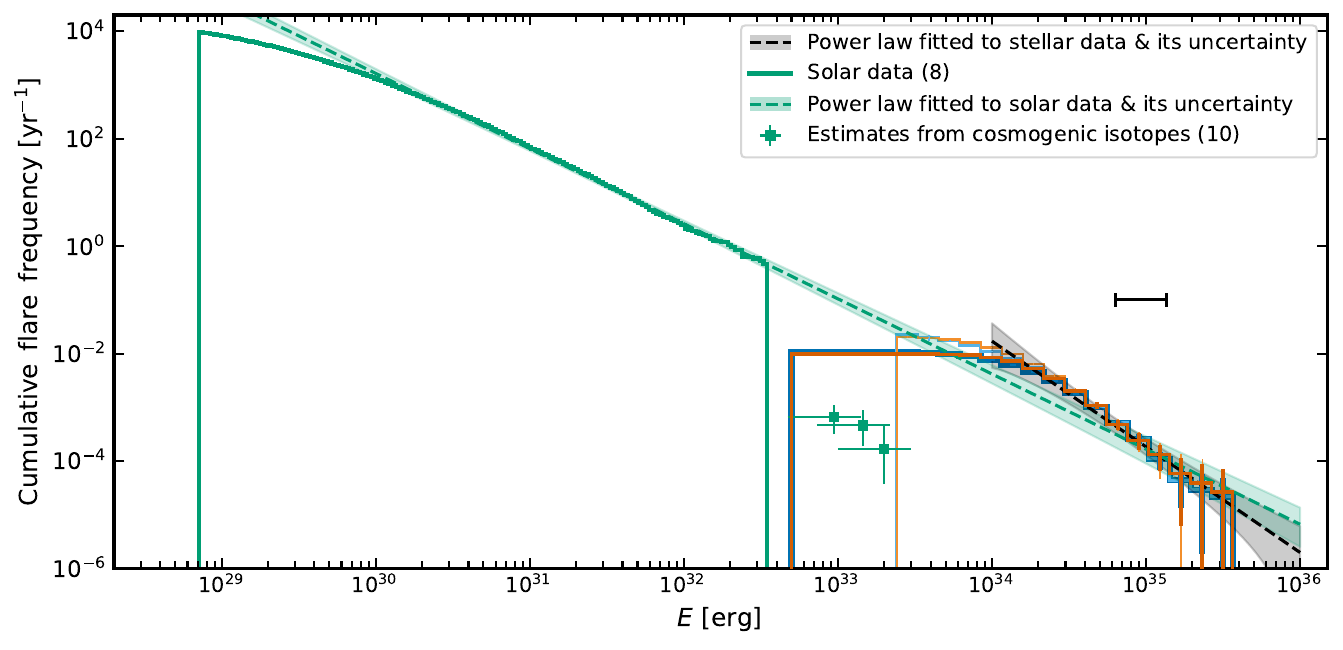}
    \caption{
    \textbf{Cumulative frequency-energy distributions of solar and stellar flares.} 
    The orange and blue histograms show cumulative data for the same samples as in Fig.~3. The black dashed line is a power law function fitted to the distribution of stellar flares with energies $>10^{34}$~erg. The green histogram shows the cumulative distribution of flares on the Sun between 1986 and 2020 \cite{Plutino2023, suppl}.
    The green dashed line is an extrapolation from
    the distribution of solar flares with energies
    $10^{30}$ to $10^{32}$~erg.   
    The green squares indicate the cumulative distribution of extreme SEP events on the Sun inferred from cosmogenic isotopes \cite{Cliver2022}. The black horizontal errorbar indicates the typical $1\sigma$ uncertainty in stellar flare energy. 
    }
\label{fig:cumul_flare_occurence_energy}
\end{figure*}

\paragraph*{Flare frequency distribution.}
We determine \cite{suppl} the flare frequency as a function of the flare energy, $E$, to quantify the number of flares per star per year per unit of energy. Figure~\ref{fig:pdf_flare_occurence_energy} shows our calculated stellar flare frequency, and Fig.~\ref{fig:cumul_flare_occurence_energy} shows its associated cumulative distribution.  Above $10^{34}$~erg the stellar superflare frequency decreases with energy, roughly following a power law, $\sim E^{-\alpha}$, where $\alpha$ is the power-law exponent. Fitting a power law model to the cumulative distribution gives $\alpha=1.97 \pm  0.30$ (Fig.~\ref{fig:cumul_flare_occurence_energy}).

Our measured frequency of stellar superflares with energies $<10^{34}$~erg is incomplete,  because their signal-to-noise ratios could fall below our detection threshold \cite{Vasilyev2022, suppl}. For the Sun, the cumulative distribution of flares also follows a power law \cite{Schrijver2012, Plutino2023}, with  $\alpha=1.399 \pm 0.056$ \cite{suppl} over a lower energy range ($10^{29}$ to $10^{33}$ erg). Although the solar and stellar measurements do not overlap in energy, an  extrapolation of the solar flare frequency distribution to higher energies is consistent with the stellar superflare frequency distribution (Fig.~\ref{fig:cumul_flare_occurence_energy}).  

The cumulative distribution of stellar superflares indicates that Sun-like stars with effective temperatures between 5000\,K and 6500\,K generate superflares with energies greater than $10^{34}$~erg with a frequency of $(8.63\pm0.20) \times 10^{-3}$ yr$^{-1}$. This stellar sample includes stars both slightly cooler and warmer than the Sun as well as stars that are currently more variable than the Sun has been at any point in the last 140 years \cite{Reinhold2020}. We quantified the stellar variability using the variability range $\Rvar$, computed as the difference between the 95th and 5th percentiles of the fluxes in a light curve, which are sorted in increasing order and normalized by their median \cite{Basri2010, Basri2011,Reinhold2020}. We found that narrowing the stellar sample to stars more similar to the Sun -- temperatures  5500 to 6000\,K and variabilities within the solar range, $\Rvar<0.3\%$ \cite{Reinhold2020} -- had very little effect on the superflare frequency (Fig.~\ref{fig:cumul_flare_occurence_energy}). Further narrowing this sample to stars with known rotational periods reduces the number of stars by almost a factor of 6 but had  minimal impact on measured flare frequency. 
Table~\ref{tab:table_1} lists our measured flare frequencies for three different sample selection cuts, along with the Sun for comparison.

\paragraph*{Comparison to previous results.}
Previous studies of stellar superflares \cite{Shibayama2013,Notsu2019,Okamoto2021} also found the flare frequency follows a power law as a function of energy. However, those studies found  frequencies approximately two orders of magnitude lower than our measurements (Fig.~\ref{fig:pdf_flare_occurence_energy}), inconsistent with an extrapolation of the solar flare distribution to higher energies. We attribute this difference to three main factors. 

Firstly, our method accounts for possible contamination by flares on background stars, cosmic rays, and minor Solar System bodies. We therefore did not need to restrict our stellar sample to isolated stars, those without nearby background sources. Previous studies \cite{Shibayama2013, Okamoto2021} excluded $\sim70\%$ of Sun-like stars from their analysis for this reason. Removing this requirement allowed us to use a larger sample size. 

Secondly, previous studies used a higher flare detection threshold in their light curve analysis to exclude background events \cite{Shibayama2013}. We expect our algorithm for subpixel localization of flares in the Kepler images to filter out those background events, so set a lower flare detection threshold of a 5$\sigma$ peak in the light curve. 

Thirdly, previous studies  \cite{Shibata2013} argued that very large starspots, with an area $> 10^{-2}$ of the stellar hemisphere, are required to generate superflares. They expected such large starspots to produce large photometric variability, sufficient to determine the  rotational period. Previous studies therefore assumed that stars with unknown rotation periods (approximately 84\% of those with Sun-like effective temperatures) are inactive and excluded them from their samples. Our analysis identified $1941$ superflares that occurred on stars with unknown rotation periods and low variability ($R_{\mathrm var} <0.3\%$)\cite{suppl}, which would have been excluded from previous studies. The low photometric variability of those stars does not imply the absence of large star spots. Photometric variability is strongly affected by the nonaxisymmetry of the surface spot distribution \cite{Emre2020, Narrett2024}, stellar inclination angles \cite{Nina2023},
as well as by bright facular features that can reduce photometric variability by compensating for spot contribution \cite{Shapiro2017, Reinhold2019}.  We discuss these differences in more detail in the Supplementary Text. This combination of factors means we measure a higher stellar superflare frequency than previous studies (Fig.~\ref{fig:pdf_flare_occurence_energy}).

\paragraph*{Implications for the Sun.}
We found that Sun-like stars produce superflares with bolometric energies  $>10^{34}$~erg roughly once per century. That is more than an order of magnitude more energetic than any solar flare recorded during the space age, about sixty years. Between 1996 and 2012 twelve solar flares had bolometric energies $>10^{32}$~erg, but none were $>10^{33}$ erg \cite{Emslie2012ApJ}. The  most powerful solar flare recorded occurred on 28 October 2003, with an estimated bolometric energy of $7 \times 10^{32}$~erg \cite{Emslie2012ApJ,Mooreetal2014}, which exceeds estimates for the Carrington Event ($4 \times 10^{32}$ to $6\times 10^{32}$~erg)  \cite{Cliver2013,Hayakawa2023}. We also found that our measured stellar superflare frequency and the solar flare distribution extrapolated to higher energies are consistent with each other, having similar power law distributions and indices $\alpha$, which might indicate a shared (super)flare generation mechanism.

The number of extreme SEP events identified in the cosmogenic isotopes record of  the last 12 millennia \cite{Usoskin2023SSRv} is substantially lower than implied by the superflare frequency of the Sun from either the extrapolation of solar flare data or our  stellar superflare measurements (Fig.~\ref{fig:cumul_flare_occurence_energy}). It is unlikely that extreme SEP events have been overlooked in cosmogenic isotope data \cite{Ilya2023}, so this inconsistency could be due to the indirect  connection between SEP events and superflares. The relationship between the energies of SEP events and superflares is highly uncertain [\!\cite{Cliver2022}, see Supplementary Text]. It is possible that superflares are rarely accompanied by extreme SEP events, as has been found for lower energy solar flares  \cite{Li2020}. 

We cannot exclude the possibility that there is an inherent difference between flaring and non-flaring stars that was not accounted for by our selection criteria. If so, the flaring stars in the Kepler observations would not be representative of the Sun. Approximately 30\% of flaring stars are known to have a binary companion \cite{Notsu2019, suppl}. Flares in those systems might  originate on the companion star or be triggered by tidal interactions. If instead our sample of Sun-like stars is representative of the Sun's future behavior, it is substantially more likely to produce a superflare than was previously thought.\\


\newpage

\begin{table*}
\caption{{\bf Measured superflare frequencies.}  
The flare frequencies derived from our measurements are listed for two energy ranges and three stellar sample selection cuts. Uncertainties are $1\sigma$. $N_\mathrm{flares}$ is the number of detected flares and $N_\mathrm{stars}$ is the number of stars they occurred on, and $P_{\rm rot}$ is the rotation peroid. For comparison, we also list values for the Sun extrapolated from solar flares at lower energies.
}
\begin{tabular}{|c|c|c|c|c|c|}
  \hline
  Sample selection & \makecell{Stars in \\sample } &
  \makecell{frequency (yr$^{-1}$) \\ 
  $10^{34} \mathrm{erg}$ to $10^{35}\mathrm{erg}$ } &
  \makecell{frequency (yr$^{-1}$) \\ $E>10^{35}\mathrm{erg}$}  & 
  $N_\mathrm{flares}$ &
   $N_\mathrm{stars}$\\
  \hline
  \makecell{G-type main sequence stars \\ 
  $5000$\,K$<T_{\rm eff}<6500$\,K \\ 
  $P_{\rm rot}>20$ d or unknown} &  \Nstarinputfinal & \makecell{$(8.63 \pm 0.20) \times 10^{-3}$} 
   &  \makecell{$(3.90 \pm 0.60)\times 10^{-4}$}  & \Nflares & \Nflaringstars\\
  \hline

  \makecell{G-type main-sequence stars \\
  $5500$\,K$<T_{\rm eff}<6000$\,K \\$\Rvar<0.3\%$ \\
  $P_{\rm rot}>20$ d or unknown} & 32,450 &  \makecell{$(9.31\pm 0.30)\times 10^{-3}$}    & \makecell{$(3.59 \pm 1.06)\times 10^{-4}$} & 1501 & 1383 \\
   \hline

  \makecell{G-type main-sequence stars \\ 
   $5500$\,K$<T_{\rm eff}<6000$\,K \\ $\Rvar<0.3\%$ \\ 
  $20$ d $<P_{\rm rot}<30$ d} & 5959 &  \makecell{$(7.70 \pm 0.95)\times 10^{-3}$} 
   &  $-$   & 238 & 208 \\
  \hline
  
   \makecell{The Sun} & - & \makecell{$(4.13^{+1.95}_{-1.38}) \times 10^{-3}$}  &  \makecell{$(1.63^{+1.18}_{-0.76}) \times 10^{-4}$} & -- & -- \\
  \hline
\end{tabular} 
\label{tab:table_1}
\end{table*}

\bibliographystyle{Science}
\bibliography{main}
\newpage
\subsection*{Acknowledgments}
We thank the three anonymous referees for constructive criticism and useful advice, which greatly improved the paper. 
Data collected by the Kepler mission was obtained from the Mikulski Archive for Space Telescopes (MAST) operated by the Space Telescope Science Institute (STScI). Funding for the Kepler mission is provided by the NASA Science Mission Directorate. STScI is operated by the Association of Universities for Research in Astronomy, Inc., under NASA contract NAS~5–26555.
Data from the European Space Agency (ESA) mission {\it Gaia} (\url{https://www.cosmos.esa.int/gaia}) were processed by the {\it Gaia} Data Processing and Analysis Consortium (DPAC, \url{https://www.cosmos.esa.int/web/gaia/dpac/consortium}). Funding for the DPAC has been provided by national institutions, in particular, the institutions participating in the {\it Gaia} Multilateral Agreement.
This research made use of \textsc{lightkurve}, a Python package for Kepler and TESS data analysis \cite{Lightkurve2018}.

\noindent
\textbf{Funding:} 
V.V. and L.G. acknowledge support from the Max Planck Society under grant “Preparations for PLATO Science” and from the German Aerospace Center (DLR) under grant “PLATO Data Center” 50OP1902. A.I.S. acknowledges funding from European Research Council Synergy Grant REVEAL 101118581. A.S.B. and L.G. acknowledge funding from European Research Council Synergy Grant WHOLE SUN 810218. A.S.B. acknowledges financial support by grants from CNES Solar Orbiter and Space Weather and CNRS-INSU/PNST. 
T.R. acknowledges support from German Aerospace Center (DLR) grant No. 50OU2101. 
I.U. acknowledges support from the Research Council of Finland (Projects 321882 and 354280).
Y. N. acknowledges support from NASA ADAP award program No. 80NSSC21K0632, Japan Society for the Promotion of Science (JSPS) KAKENHI grant No. 21J00106, and the International Space Science Institute, which supports International Team 510, "Solar Extreme Events: Setting Up a Paradigm.”
H.M. was supported by JSPS KAKENHI Grant Numbers JP20H05643, JP21H01131, JP24K00685, JP24H00248, and 24K00680. 

\noindent
\textbf{Author contributions:}  V.V., A.I.S., S.K.S. and I.U. conceived the study. A.I.S., S.K.S., N.A.K. supervised the project. V.V. and T.R. analyzed the Kepler and Gaia data and obtained the final results. V.V. wrote the software. 
All authors contributed to the methodology, interpretation, and manuscript preparation.

\noindent
\textbf{Competing interests:} There are no competing interests to declare.

\noindent
 \textbf{Data and materials availability:} 
 The Kepler Light curves and images (target pixel files) are available from MAST at \url{https://archive.stsci.edu/kepler/data_search/search.php} or using the \textsc{lightkurve} package \cite{Lightkurve2018}, in either case using the Kepler IDs listed in Data S1. The Gaia data are available through the ESA Gaia Archive  at \url{https://gea.esac.esa.int/archive/} by selecting 'single object' then using the Gaia DR3 IDs in Data S1.
 The stars we selected for our sample are listed in Data S1, and the flares we identified are listed in Data S2. 
The flare detection code \cite{Vasilyev2022} is  available at \url{https://github.com/ValeriyVasilyevAstro/LOSE} and archived at Zenodo \cite{vasilyev_2024_13951243}.

\noindent
\textbf{Supplementary Materials:}\\
Material and Methods \\
Supplementary Text \\
Figures~\ref{fig:Figure9} to \ref{fig:RUWE} \\
Table ~\ref{Table_3:fast_rot_tab}, 
References \textit{(45-67)} \\
Data~S1 and S2\\
\noindent
\newpage
\renewcommand{\thefigure}{S\arabic{figure}}
\setcounter{figure}{0}
\setcounter{page}{1}

\renewcommand{\thetable}{S\arabic{table}}
\setcounter{table}{0}
\setcounter{page}{1}
\begin{figure}[t]
  \centering
  \includegraphics[width=0.3\textwidth]{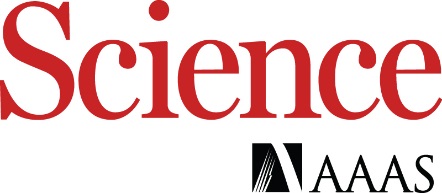}
\end{figure}
{
\centering
\Large
{Supplementary Materials for} \\[\baselineskip]
\textbf{Sun-like stars produce superflares roughly once per century} \\[\baselineskip]
\normalsize

Valeriy Vasilyev \textit{et al}. \\[\baselineskip]

\small
$^\ast$ Corresponding author: Valeriy Vasilyev, vasilyev@mps.mpg.de \\[\baselineskip]
}

\begin{flushleft}
\textbf{This PDF file includes:} \\
\hspace{1.0cm} Materials and Methods \\
\hspace{1.0cm} Supplementary Text \\
\hspace{1.0cm} Figures~S1 to S7 \\
\hspace{1.0cm} Table~S1 \\
\hspace{1.0cm} References (45-67) \\
\textbf{Other Supplementary Materials for this manuscript include the following:} \\
\hspace{1.0cm} Data~S1 and S2 (machine-readable data tables)
\end{flushleft}{}

\newcounter{defcounter}
\setcounter{defcounter}{0}
\newenvironment{myequation}{%
\addtocounter{equation}{-1}
\refstepcounter{defcounter}
\renewcommand\theequation{S\thedefcounter}
\begin{equation}}
{\end{equation}}

\newpage
\renewcommand{\theequation}{S\arabic{equation}}
\noindent
\paragraph*{Materials and Methods}
\paragraph*{Sample selection.}  
To obtain the initial sample of potentially flaring stars, we selected 76,418 FGK-type stars with effective temperatures from 5000 to 6500$\,$K according to stellar parameter catalogs \cite{Mathur2017,Berger2020} and absolute magnitudes $4<M_{\rm G}<6$ mag (computed from apparent magnitudes, distances, and extinctions) from the Gaia data release 3 (DR3) \cite{GaiaDR3}. We removed known eclipsing binaries \cite{Kirk2016} and constrained our sample to stars with known radii \cite{Berger2020}. To focus on stars with properties similar to the Sun, we removed stars with rotation periods less than 20 days using the rotation period catalog \cite{Reinhold2023}. 
In most cases we used the final rotation period listed in that catalog, but in some cases we adopted the auto-correlation function period instead (see below) \cite{Reinhold2023}.

To restrict the sample to single stars with constrained astrometry, we removed stars with Renormalized Unit Weight Error (RUWE) greater than 1.4 \cite{LindegrenGaia2018}  listed in Gaia DR3.  $\rm RUWE>1.4$ indicates that Gaia's single-star model provides a poor fit to the astrometric observations, implying that the source is either non-single or otherwise problematic for the astrometric solution. RUWE is an indicator of multiplicity \cite{Belokurov2020,Stassun2021}. We expect this RUWE cut to remove unresolved systems with semi-major axes between 0.1 and 10 astronomical units for stars with distances 1 to 2 kpc from the Sun \cite{Belokurov2020}. After completing our analysis, we confirmed that all the flaring stars detected by our algorithm have distances $<2$ kpc, with a median of 1 kpc.

Applying these selection criteria left us a sample of \NstarinputBeforeFastRotatorsCleaning stars, to which we applied our flare detection algorithm. We re-determined the rotation periods of all stars flaring multiple times. 
\\

\noindent
\paragraph*{Kepler data.} The archival Kepler light curves had already been corrected for instrumental systematics \cite{Stumpe2012,Smith2012,Stumpe2014}, including removal of shared signals across the detector from the simple aperture photometry (SAP) light curves. We use the presearch data conditioning - SAP (PDC-SAP) flux from the Kepler Data Release 25 \cite{KeplerDr25}, for which most instrumental effects had already been removed. 
We analyze the long-cadence data with an exposure time of $\sim 30$ minutes, using the \textsc{lightkurve} package \cite{Lightkurve2018} to retrieve the data.\\

\noindent
\paragraph*{Flare detection.} 
The flare detection procedure consists of two steps: i) a search for potential flare events in the light curves, and ii) localization of these events on the detector. For the first step, in each Kepler data segment we apply a moving average filter with a width of 15 data points ($\sim7.5$ hours) to produce a smoothed light curve. Then, we search for data points in the light curve that are more than   $5\sigma$ above the smoothed light curve for at least two subsequent data points (as occurs for the example flare in Fig.~\ref{fig:Figure1}). 

In the second step, we examine the images (the target pixel files) for all potential flare events satisfying the $5\sigma$ criterion. This time series of 2-dimensional images contains information on how the stellar photon flux is distributed across the detector. We localize the flare on the detector taking into account the point response function of the Kepler instrument \cite{Vasilyev2022}. For the two images corresponding to the two highest consecutive data points above the 5$\sigma$ threshold (separated by $\sim30$ min in time), we subtracted an image with the quiet stellar flux (the stellar flux in absence of a flare) to isolate light from the flare. To obtain the image of the quiet stellar flux (during the flare), we selected images from 0.6 days before and after the flare (excluding those affected by the flare itself). We model the time evolution of the quiet stellar flux in each pixel in images using cubic polynomial and use interpolation to predict the quiet stellar flux during the flare.

To determine the most probable location of the flux excess on the detector, we fitted the instrumental point spread function to those two images separately. We use a flare detection method \cite{Vasilyev2022} based on the Markov Chain Monte Carlo ensemble sampling to compute the marginalized probability for the flare localization. From the probability distribution functions we compute the flare location falling into the 68\%, 95\%, and 99.9\% confidence level ellipses. 
Then we compare their location with the position of the target star by converting star's right ascension and declination listed in the Gaia DR3 catalog \cite{GaiaDR3} to the instrument pixel coordinates. 

Figure~\ref{fig:Figure9} shows the possible outcomes from our flare localization process. In cases where the two confidence ellipses intersect, the potential source of the flux excess for both images is situated in the same location. We consider five distinct cases:
\begin{enumerate}[label=(\Alph*)]
\item The area $S$ withing the intersection of the two 99.9\% confidence ellipses does not contain any known target stars. We discard these events.
\item A known background star (listed in the Gaia DR3 catalog) lies within $S$, but the target star does not. We conclude that the flare is associated with the background star and discard such cases from the analysis.
\item The target star lies within the 99.9\% confidence ellipse for both cadences, and no known background star does. We conclude that the flare is associated with the target star and use it in our analysis.
\item The intersecting area $S$ contains both the target and exactly one known background star. We cannot distinguish between flares on the target or on the background star and discard the event from our analysis.
\item The intersecting area $S$ contains more than one known background star, but the target star does not. We discard these events.
\end{enumerate}

We expect any observed flare on a specific star to occur without any influence from the timing of flares on other stars, which would produce a flat distribution of flares as a function of time. However, the presence of residual instrumental effects, such as a sudden increase in instrumental noise within a short time period, can generate outliers that can be mistakenly interpreted as flares \cite{Vasilyev2022}. To remove contamination of our flare sample from residual instrumental effects, we analyze the temporal distribution of flares across the Kepler observation window (Fig.~\ref{fig:s1_time_distir}). We identified 11 days in time with the number of flares per day more than $5\sigma$ above normal. We exclude all \NumberOfFlaresInDaysWithTooLargeFlareRate events that occurred on those days from our analysis.

\noindent
\paragraph*{Rotation periods.} \label{sec:fast_rotators} 
We adopt rotation periods from a published rotation period catalog \cite{Reinhold2023}. The final periods $P_{\rm rot, \,fin}$ listed in that catalog were determined with two different methods, auto-correlation function (ACF) and gradient of the power spectrum (GPS) \cite{Reinhold2023}. Hereafter, we denote rotation periods measured by ACF as $P_{\rm rot, \, ACF}$ and  those measured by GPS as $P_{\rm rot, \, GPS}$. In some cases the GPS method finds twice the period returned by the ACF method. Our visual inspection of the Kepler light curves showed that in the majority of cases the ACF period is more likely to be correct, especially for cases with large photometric variability quantified by $\Rvar$. In cases where $0.4<P_{\rm rot, \, ACF}/P_{\rm rot, \, GPS} < 0.6$, ACF local peak height $>0.3$, and mean variability $\Rvar>0.3\%$, we adopted the GPS by the ACF periods.

Before analyzing the flare occurrence rate, we cleaned our sample by removing fast rotators defined as all stars with adopted rotation periods $\Prot<20$ days. These stars are likely younger and therefore more active that the Sun, resulting in a higher flare occurrence rate than slower rotating Sun-like stars.  

 The resulting flare sample contained 335 stars flaring multiple times, some of them dozens of times. 
 Since the flaring rate increases with faster rotation [\!\cite{Okamoto2021}, see their figure~10], such a high flaring rate is uncommon for slowly rotating Sun-like stars, and might indicate that these flares originate from a fast-rotating star falling in the same aperture as the target star (Fig.~\ref{fig:fast_rotators}) or that the rotation period was not determined correctly.  So we checked all stars flaring multiple times if their rotation period was determined correctly and was greater then 20 days. To remove these contaminants, we computed the generalized Lomb-Scargle periodogram (GLS \cite{Zechmeister2009}) for all flaring stars and excluded those with significant peaks above a common threshold of 0.1 \cite{Reinhold2015} for all periods lower than 20 days. This removed stars flaring multiple times but there were still 242  cases left. We inspected those light curves by eye, computed the ACF, and found several cases with short periods that were not the highest peak detected by the GLS. This process identified \nfast stars with periods shorter than 20 days, which are listed in Table~\ref{Table_3:fast_rot_tab}. Of these \nfast stars, \nfastper had a final period reported in the rotation period catalog \cite{Reinhold2023}, in most cases with periods longer than 20 days. 7 of the remaining 9 stars were not listed in the catalog in \cite{Reinhold2023} because they have the surface gravity $\rm log\,g<4.0$, and 2 stars did not fulfill minimum ACF peak height criteria.  These 46 stars show 547 flares over the four years of observations. We have excluded them
from subsequent analysis.

To visualize the difficulties in determining the correct rotation period, we show the example of the star KIC\,7772296 in Fig.~\ref{fig:fast_rotators}. 
The highest peak is usually the most likely rotation period, at least for periodic signals. However, additional signals in the time series alter the shape of the ACF and make it more difficult to determine the correct rotation period, sometimes leading to false peaks in an automated period search. Long-term signals could be caused by a superposition of two stars within the same aperture, or by residual instrumental effects, which have not been removed from the time series. For  KIC\,7772296, a rotation period of 23.34 days was reported in the catalog \cite{Reinhold2023}, but the shape of the ACF indicates a superposition of at least two periods. We therefore identify the shorter period of 1.19 days as the rotation of the flaring star. The Gaia DR3 data does not indicate this star consists of two sources, so we suspect that KIC\,7772296 might be an unresolved binary.

After excluding these 46 stars, our final sample consists of \Nflares flares on \Nflaringstars stars.
\\

\noindent
\paragraph*{Flare energy.} The flare energy is calculated as the time integral of the bolometric flare luminosity:
\begin{equation} \label{eq:flare_energy1}
E = \int L_\mathrm{flare}(t) dt,
\end{equation}
where $t$ is time and $L_\mathrm{flare}(t)$ flare luminosity.
We perform the integration in a time window starting from the first point in the light curve that is equal to or above the $5\sigma$ level and ending at the descending phase when the flare flux crosses the $1\sigma$ level again, determined by linear interpolation between the last data point above and the first below the $1\sigma$ level.  

We assume that a white-light flare  (as seen in the visible wavelength band) is a black body radiator with an effective temperature $T_\mathrm{flare}=9000 \pm 1000$\,K \cite{Guenther2020}, so its bolometric luminosity is 
\begin{equation} \label{eq:flare_energy2}
L_{\rm flare}(t) = A_{\rm flare}(t) \sigma_\mathrm{SB} T_{\rm flare}^4,
\end{equation}
where $A_\mathrm{flare}(t)$ is the area of the flare projected onto the sky, and $\sigma_\mathrm{SB}$ is the Stefan-Boltzmann constant. 

The number of counts in each pixel of the charge-coupled device (CCD) detector, with a known transmission function $\Phi (\lambda) $ (where $\lambda$ is the wavelength), is approximately proportional to the number of photons hitting the pixel during the exposure time.  Therefore, the photon luminosities of the flare, $L'_\mathrm{flare}$, and the star, $L'_\mathrm{*}$, in the passband are given by
\begin{equation} \label{eq:flare_energy3}
L'_\mathrm{flare}(t) =  A_\mathrm{flare}(t) \int \Phi (\lambda) B_\lambda (T_\mathrm{flare})   (hc/\lambda)^{-1} d\lambda
\end{equation}
and 
\begin{equation} \label{eq:flare_energy4}
L'_\mathrm{*} =  \pi R_*^2 \int \Phi (\lambda) B_\lambda (T_*) (hc/\lambda) ^{-1}d\lambda,
\end{equation}
where $B_\lambda$ is the Planck function, $h$ the Planck constant,  $c$ the speed of light, $T_*$ the stellar effective temperature and $R_*$ the stellar radius. The observed flux excess caused by the flare in the normalized light curve is related to the flare's and star's photon luminosities in the passband:
\begin{equation} \label{eq:flare_energy5}
\Big(\delta l/l\Big) (t) = L'_\mathrm{flare}(t)/L'_\mathrm{*},
\end{equation} 
where $l(t)$ is the averaged flux and $\delta l (t)$ is the flux excess caused by the flare. Combining equations~(\ref{eq:flare_energy3} to \ref{eq:flare_energy5}), we determine the flare area:
\begin{equation} \label{eq:flare_energy6}
 A_\mathrm{flare}(t) = \Big(\delta l/l\Big) (t) \pi R_*^2  \frac{ \int \Phi (\lambda) B_\lambda (T_*)  \lambda  d\lambda} {\int \Phi (\lambda) B_\lambda (T_\mathrm{flare}) \lambda d\lambda}.
\end{equation}
Combining Eq.~\ref{eq:flare_energy1}, \ref{eq:flare_energy2}, and ~\ref{eq:flare_energy6} we calculated the flare energy:
\begin{equation} \label{eq:flare_energy}
E =   \sigma_\mathrm{SB} T_{\rm flare}^4   
\pi R_*^2  \frac{ \int \Phi (\lambda) B_\lambda (T_*)  \lambda  d\lambda} {\int \Phi (\lambda) B_\lambda (T_\mathrm{flare}) \lambda d\lambda}
\int       
\Big(\delta l/l\Big) (t)
dt.
\end{equation}
The stellar parameters,  $R_*$ and $T_*$, are taken from a previously published Gaia-Kepler stellar properties catalog \cite{Berger2020}.  

The uncertainty of the flare energy is determined from the uncertainties in (i) the flux in the light curve ($1\sigma$ scatter), (ii) the stellar effective temperature,  (iii) the stellar radius, and (iv) the flare temperature. For each flare, we compute 1000 realizations of the flare energy, assuming that each of these five parameters is Gaussian distributed around the given value with the standard deviation equal to the uncertainty of the parameter. The resulting scatter of the individual flare energy distribution (i.e., the standard deviation) is taken as the uncertainty. For all \Nflares flare energies, the median uncertainty is around 37\% (Fig.~\ref{fig:pdf_flare_occurence_energy}). 

Fig.~\ref{fig:Energies_vs_teff} shows the measured flare energies as a function of stellar effective temperatures. In the light curve analysis, we detect superflares whose flux exceeds the $5\sigma$ detection threshold. i.e. flares below the threshold are not identified by our detection algorithm. However, the same relative change of the stellar brightness ($\delta l/l$) (due to a superflare) corresponds to a higher flare energy for hotter stars and a lower energy for cooler stars. Therefore, the minimum superflare energy that our method can detect is higher for hotter stars. For example, most of the flares with energies $<10^{34}$~erg on stars with $T_\mathrm{eff}>6000$\,K fall below the detection threshold. We discuss the effect of this bias below.  

A previous study found that the flare energies based on the long-cadence Kepler data (utilized in this study) are systematically underestimated by about 25\% \cite{Yang2018ApJ}. We accounted for this effect by  dividing all flare energies by a factor of 0.75.
\\

\noindent
\paragraph*{Solar flare data.} 
Solar flares are continuously observed in the soft X-ray (SXR) using space-based instruments \cite{Veronig2002, Aschwanden2012ApJ, Plutino2023}. We used a catalog of 334,122 solar flares detected in SXR observations by the Geostationary Operational Environmental Satellite (GOES)  satellite constellation from 1986 to 2020 \cite{Plutino2023}. We use a previously determined  relationship between the total radiated energy of the flare, $E$,  and the SXR peak flux, $C_\mathrm{GOES}$ \cite{Cliver2022}:
\begin{equation}
E = 0.33 \times 10^{32} \Bigg( \frac{\mathcal{C_\mathrm{GOES}}}{\mathcal{C}_\mathrm{GOES, X1}}   \Bigg)^{0.72}, 
\end{equation}
where $\mathcal{C}_\mathrm{GOES, X1}=10^{-4}$ W m$^{-2}$ is the peak SXR flux of an X1 class flare  used for normalization. The resulting solar flare energies are shown in Fig.~\ref{fig:cumul_flare_occurence_energy}. Fitting a power law to this distribution gives an exponent of
$\alpha = 1.399 \pm 0.056.$.

We also considered other catalogs of solar flares observed in extreme ultraviolet (EUV) \cite{Aschwanden2000ApJ} and hard X-rays (HXR) \cite{Crosby1993}.  However, we were unable to directly compare these to stellar flare energies due to the absence of an established method for converting EUV and HXR peak fluxes to bolometric fluxes and energy. We therefore chose to restrict our analysis to the SXR data \cite{Cliver2022}.
\\

\noindent
\paragraph*{Flare frequency as a function of flare energy.} 
We denote $n_\mathrm{flares}(E)$ as the number of flares on all analyzed stars with energies larger than $E$ and $\mathrm{d} n_\mathrm{flares}(E)$ as the number of flares with energies between $E$ and $E+ \mathrm{d} E$. The flare frequency, $f(E)$,  characterises the number of flares per star per year per unit energy and is given by
\begin{equation}
f(E) = \frac{ \mathrm{d} n_\mathrm{flares} (E)}{ n_\mathrm{stars} \Delta t \mathrm{d} E}, 
\end{equation}
where $\Delta t$ is the duration of observations, and $n_\mathrm{stars}$ is the total number of observed stars.

The cumulative flare frequency distribution, $F(E)$, the number of flares per star with energy greater than $E$, is given by:
\begin{equation}
F(E) = \int_{E}^{\infty} f(E') \mathrm{d} E' =  \frac{ n_\mathrm{flares}(E)}{ n_\mathrm{stars} \Delta t}.
\end{equation} 

In each energy bin, the uncertainty of the cumulative energy distribution, $\Delta F$, is derived from the uncertainties in the number of stars, $\Delta n_\mathrm{stars}$, and the number of flares in the bin,  $\Delta n_\mathrm{flares}$. We use standard error propagation to estimate the uncertainty $\Delta F$: 
\begin{equation} \label{eq:delta_f}
\Delta F = \frac{1}{\Delta t} \sqrt{\Bigg(  \frac{\Delta n_\mathrm{flares}  }{  n_\mathrm{stars}}  \Bigg) ^2   +  \Bigg( \frac{\Delta n_\mathrm{stars} n_\mathrm{flares} }{  n_\mathrm{stars}^2}  \Bigg) ^2    }, 
\end{equation}
where  $\Delta n_\mathrm{stars}=\sqrt{n_\mathrm{stars}}$,  $\Delta n_\mathrm{flares}=\sqrt{ 
 n_\mathrm{flares} + \delta \mathcal{E}^2 }$, and $\delta \mathcal{E}$ denotes an additional uncertainty in the number of flares in the given energy bin due to uncertainties in the individual flare energy estimates. We estimate $\delta \mathcal{E}$ from Monte Carlo simulations. Assuming a Gaussian distribution for each flare's energy, with the standard deviation equal to the uncertainty in the flare energy, we generate 1000 realizations of all \Nflares flare energies. Then, while keeping the same energy bins, we estimate how the number of flares in each energy bin changes. 
Finally, using Eq.~\ref{eq:delta_f} we compute the uncertainty in the cumulative energy distribution, $\Delta F$.
 \\

\noindent
\paragraph*{The detection threshold's effect on frequency.} \label{flare_detectability}
Our flare detection algorithm does not detect flares with low energies. To quantify the fraction of detectable flares at a given energy (at fixed effective temperatures), we performed Monte-Carlo simulations. From the sample of \Nstarinputfinal stars, we arbitrarily selected 10,000 light curves with a duration of 3 months. Then we averaged the temporal profile of all detected flares (Fig.~\ref{fig:average_flare}) and injected 25 flares with energies ranging from $5\cdot 10^{32}$~erg to $5 \cdot 10^{35}$~erg into each light curve. We then ran these synthetic light curves through our detection algorithm. For each flare energy and stellar effective temperature, we computed the ratio of detected flares to the total number of injected flares, $r(E, T_\mathrm{eff})$. 
In each flare energy bin, we estimate the true number of flares, $\mathrm{d} {n'}_\mathrm{flares}$,  from the observed number of flares, 
$\mathrm{d}n_\mathrm{flares}$  as 
\begin{equation} \label{eq:correction}
\mathrm{d} {n'}_\mathrm{flares} (E) = \sum_{\gamma} \mathrm{d} n_\mathrm{flares}(E, T_{\mathrm{eff}, \gamma})/ r(E, T_{\mathrm{eff}, \gamma}),
\end{equation}
where the summation is performed over all stellar effective temperature bins, and $\gamma$ is the temperature bin index. The uncertainty in the true number of flares is calculated as 
\begin{equation}
{\sigma'}(E) = \sqrt{ \sum_{\gamma}  \Bigg [  \frac{  \mathrm{d} n_\mathrm{flares}(E, T_{\mathrm{eff}, \gamma}) }{ r^2(E, T_{\mathrm{eff}, \gamma}) } +   \frac{\mathrm{d} n^2_\mathrm{flares}(E, T_{\mathrm{eff}, \gamma})  \mathrm{d} r^2(E, T_{\mathrm{eff}, \gamma}) } {r^4(E, T_{\mathrm{eff}, \gamma})  }  \Bigg ]    }
\end{equation}

Fig.~\ref{fig:detectability_Energies_Teff} shows  $r$ as a function of effective temperature and flare energy. For mid- to low-energy flares, the fraction of detected flares decreases with increasing effective temperature, in qualitative agreement with Fig.~\ref{fig:Energies_vs_teff}. Flares with energies $E>5 \cdot 10^{34}$~erg are detected on all stars ($r=1$) with temperatures between 5000\,K and 6500\,K. For energies $E<2 \cdot 10^{33}$~erg and effective temperatures $T_\mathrm{eff}>6000$\,K, flares are not detected, i.e. $r=0$. Therefore, we used  Eq.~\ref{eq:correction} to correct the number of detected flares with energies larger than $2 \cdot 10^{33}$~erg and  recomputed the flare frequency and the cumulative distribution for the whole sample and the solar sub-sample (Fig.~\ref{fig:pdf_flare_occurence_energy} and \ref{fig:cumul_flare_occurence_energy}).\\

\noindent
\paragraph*{Contamination by background sources.} 
We investigate contamination of our flare detections by unresolved background sources within the photometric aperture of the target star. Potential background sources include fast-rotating K- or M-dwarf stars, which are  much more active than Sun-like stars.  They have a  higher flare frequency and often larger stellar brightness increases during flares compared to the quiescent flux, by up to a factor of 16 \cite{Chang2018,Guenther2020}.

During the flare localization described above, we determined the intersecting area $S$ of the two 99.9\% confidence ellipses inside the total aperture size $Q$, which defines the CCD pixels used to extract the flux of the target star. If a background event happens within the aperture, the probability to find it within the intersecting area $S$ is $p=S/Q$ and to find it outside of $S$ is $1-p$.

For simplicity we consider consider the case when the photometric aperture contains only the target star, i.e. no background stars listed in the Gaia DR3 catalog. The number of such cases (each consisting of at least two consecutive time steps with flux above the $5\sigma$ threshold) is $13,748$, which we label as  $N^{(\mathrm{obs})}$. Among these, there are 
$9784$ (hereafter, $ N_{\mathrm{out}}^{(\mathrm{obs})}$) events where the target star is located outside of $S$ and  $3964$ (hereafter, $N_\mathrm{flare}^{(\mathrm{obs})}$) events where the target star is located within $S$.

%

For each of the $N^{(\mathrm{obs})}$ cases, we have a photometric aperture $Q$ and an intersecting area $S$. We examined the $S/Q$ distribution for all $N^{(\mathrm{obs})}$ events and calculated a median value of $p = 0.07$, which we use in our calculations.

We assume that unknown background sources are uniformly distributed throughout the entire Kepler field of view, and so also over the aperture $Q$. We denote $N^{(\mathrm{true})}_{\mathrm{bkg}}$ as the total number of events on unknown background sources in $N^{(\mathrm{obs})}$  and $N^{(\mathrm{true})}_{\mathrm{flare}}$ as the total number of true flares on the target star in $N^{(\mathrm{obs})}$. Then, the observed number of events $N^{(\mathrm{obs})}$ is 
\begin{equation}
N^{(\mathrm{obs})} = N^{(\mathrm{true})}_{\mathrm{bkg}} + N^{(\mathrm{true})}_{\mathrm{flare}}.
\end{equation}
Some fraction of these events on unknown background sources (by projection) contains the target star in $S$ and, therefore, contaminates the observed number of flares on the target star:
\begin{equation} \label{eq:target}
N_\mathrm{flare}^{(\mathrm{obs})} = N^{(\mathrm{true})}_{\mathrm{flare}} + p N^{(\mathrm{true})}_{\mathrm{bkg}}
\end{equation}
Then, the number of events on unknown background sources, when the target lies outside of $S$ is 
\begin{equation} \label{eq:background}
N_{\mathrm{out}}^{(\mathrm{obs})}  = (1 - p)
N^{(\mathrm{true})}_{\mathrm{bkg}}.
\end{equation}

Combining Eq.~\ref{eq:target} and \ref{eq:background} we obtain the true fraction of flares on the target star in the observed flare sample:
\begin{equation} 
\frac{N^{(\mathrm{true})}_{\mathrm{flare}}}{N_{\mathrm{flare}}^{(\mathrm{obs})}}  = 1 - \frac{ N_{\mathrm{out}}^{(\mathrm{obs})} } {N_\mathrm{flare}^{(\mathrm{obs})}} \frac{p}{(1 - p)}.
\end{equation}
Inserting our measured values, we find that approximately $1 - \frac{N^{(\mathrm{true})}_{\mathrm{flare}}}{N_{\mathrm{flare}}^{(\mathrm{obs})}} \approx20$\% of the observed flares we associate with Sun-like stars are likely to be events from unknown background sources. Assuming that this pollution is independent of flare energy, we apply a correction for unknown background sources by multiplying our measured flare frequency distributions by a factor 0.8.

\noindent
\paragraph*{Contamination by binaries.} 
Approximately 30\%
of flaring stars are spectroscopic binaries \cite{Notsu2019}. Our detected flares could therefore be contaminated by flares associated with close K- and M-dwarf companions of the target star. To account for this, we exlcuded stars with $\rm RUWE>1.4$ in our sample selection (see above).

To test the effectiveness of this cut, Figure~\ref{fig:RUWE} shows the distributions of RUWE for all flaring and non-flaring stars. We find the  distributions are almost identical. The mean RUWE value of flaring stars is $1.019\pm0.071$, which for non-flaring stars it is $1.015\pm0.073$. We therefore conclude that any contamination by binaries is expected to be similar for flaring and non-flaring stars.


The RUWE criterion is sensitive to unresolved binaries \cite{Belokurov2020}, but it is less sensitive to binaries with semi-major axis below 0.1 astronomical units. We, thus, cannot fully exclude contamination by more active K- and M-dwarf companions which do not manifest themselves in photometric variability because their flux is diluted by the main star. Such contamination is unlikely, as active K- and M-dwarfs usually experience multiple flares during the Kepler observation period \cite{Davenport2016, Guenther2020}. In our flaring sample, however, only 2 to 3\% of the flaring stars flare more than twice.


\begin{figure*}
    \centering
    \includegraphics[width=1.0\textwidth]{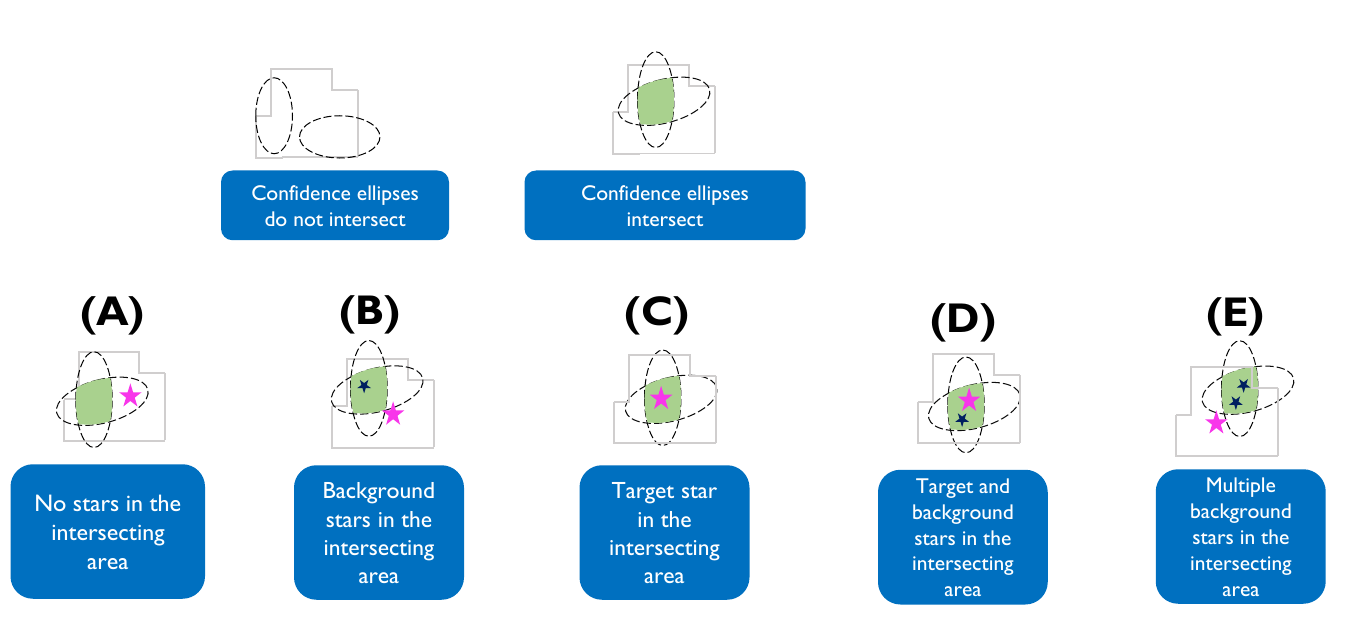}
    \caption{ {\bf Classification of results of the flare localization on images.} We associate the flare with a specific target star (purple star symbol) if it lies whithin the 99.9\% confidence ellipses (dashed ellipses) for the probable location of the flare for both images. i.e  in the intersecting area of ellipses (green shading). The light gray polygon  represents the aperture around the target star, i.e., the pixels on the image used to extract the flux of the target star. 
    }
    \label{fig:Figure9}
\end{figure*}

\begin{figure*}
    \centering
    \includegraphics[width=1.0\textwidth]{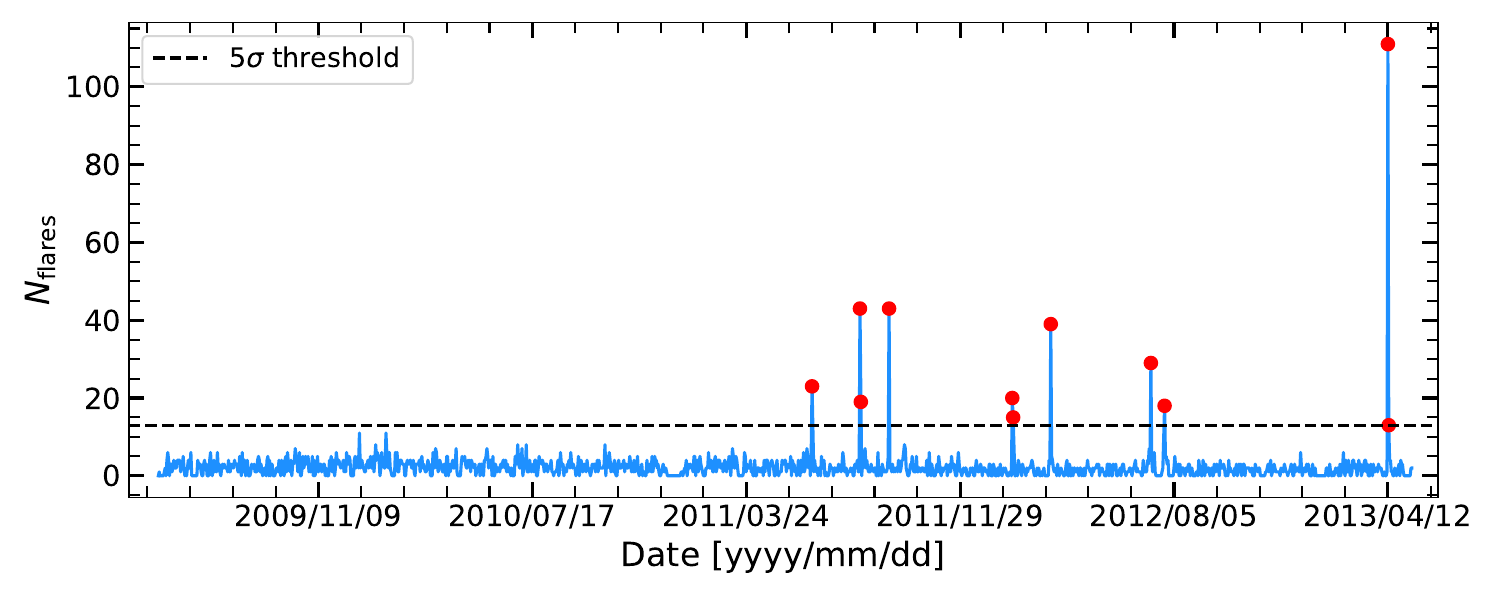}
    \caption{ {\bf Temporal distribution of all identified flares during the entire 4-year Kepler observational campaign.} 
    Red points indicate 11 peaks with a high number of events per day, exceeding the normal level by $5\sigma$ (black dashed line). 
    The data are binned with a 1-day bin width. 
    }
    \label{fig:s1_time_distir}
\end{figure*}

\begin{figure*}
    \centering
    \includegraphics[width=0.95\textwidth]{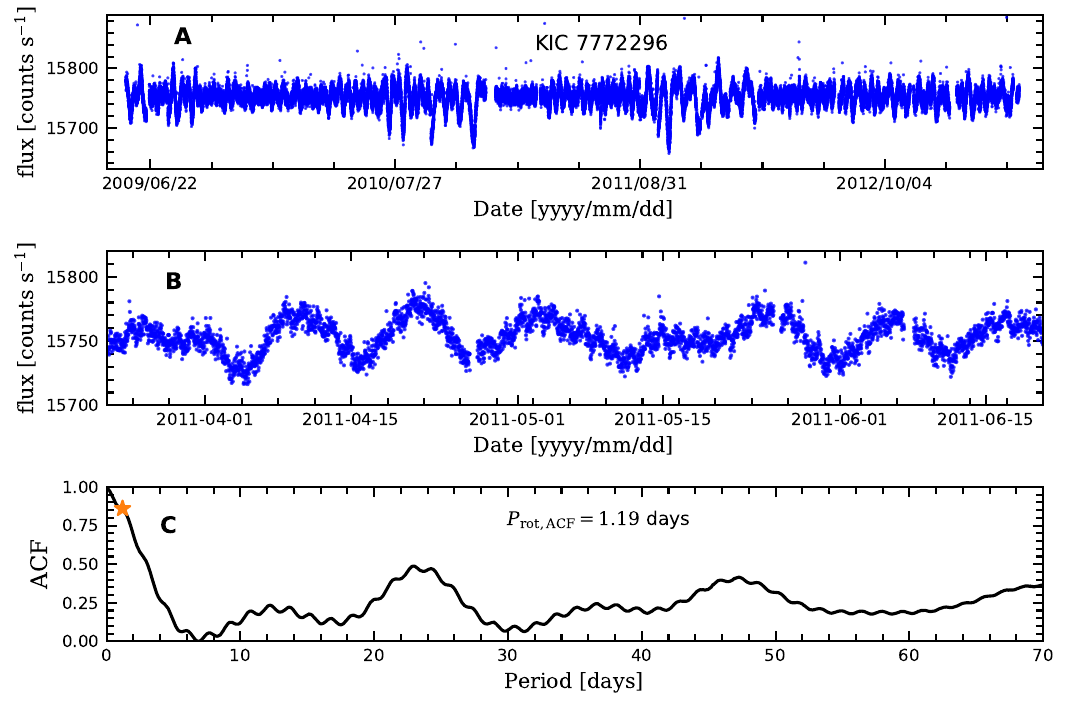}
    \caption{{\bf Example of a star excluded because it is a fast rotator.}
    \textbf{(A)} The full 4-year time series of the star KIC\,7772296. \textbf{(B)} A 90-day zoom into the light curve showing a superposition of long- and short-term periodicity. \textbf{(C)} The auto-correlation function (ACF), showing a pronounced peak at a period of 23.34 days, superimposed by a short-term periodicity. The shorter period at 1.19 days is indicated by the orange star symbol.}
    \label{fig:fast_rotators}
\end{figure*}

\begin{figure*}
    \centering
    \includegraphics[width=1.0\textwidth]{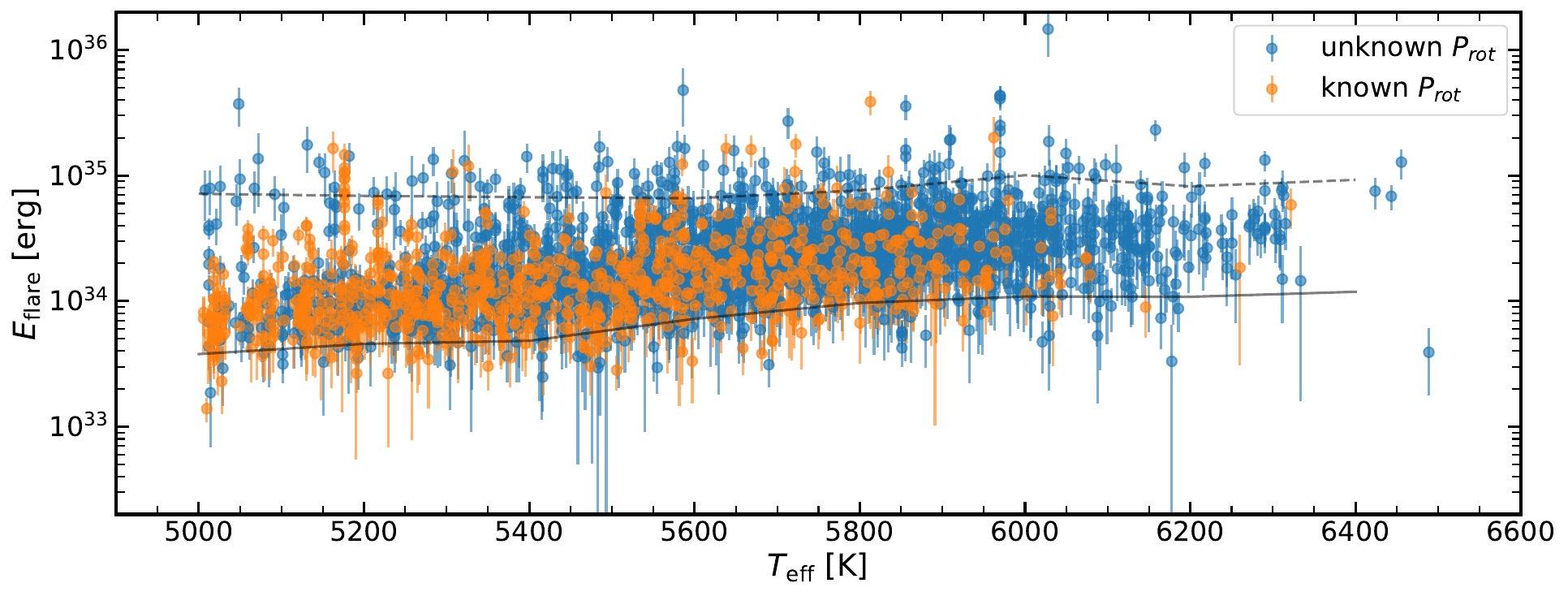}
    \caption{ {\bf Flare energy as a function of the effective temperature of the target star.} Blue and orange dots show stars with known and unknown rotation periods, respectively. The solid and dashed grey curves indicate the 5th and 95th percentiles, respectively. Vertical error bars indicate the $1\sigma$ uncertainty in measured flare energies.}
    \label{fig:Energies_vs_teff}
\end{figure*}

\begin{figure*}
    \centering
\includegraphics[width=0.5\textwidth]{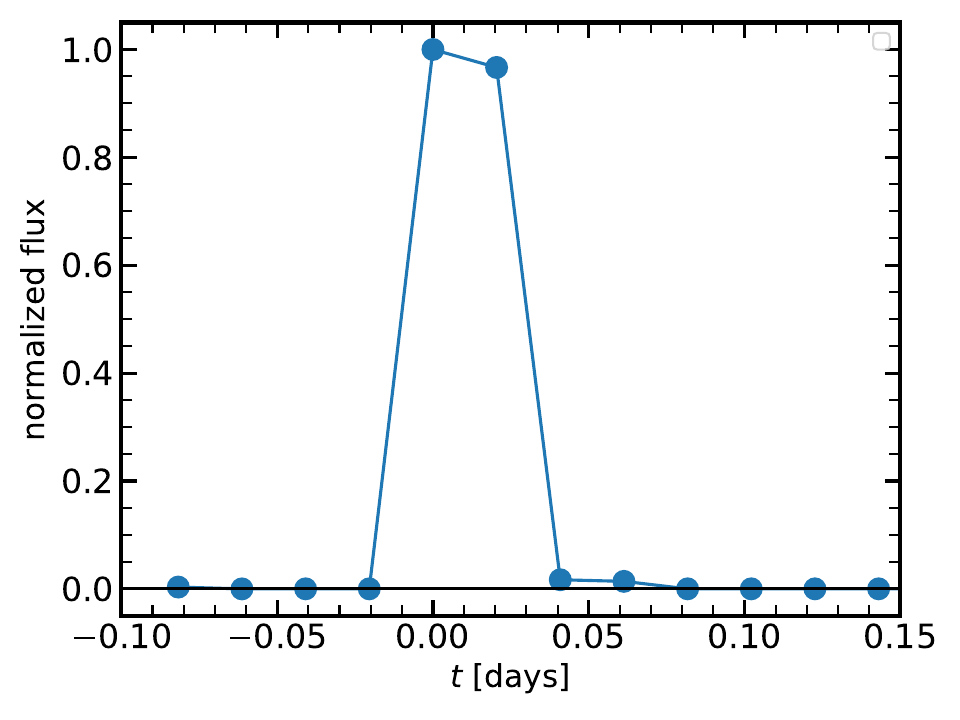}
    \caption{ {\bf Average flare profile.} Each flare was shifted in time to start at time $t=0$ and normalized by the flux at the start time. Each point in the averaged profile represents the median flux over all identified flares.}
    \label{fig:average_flare}
\end{figure*}

\begin{figure*}
    \centering
    \includegraphics[width=1.0\textwidth]{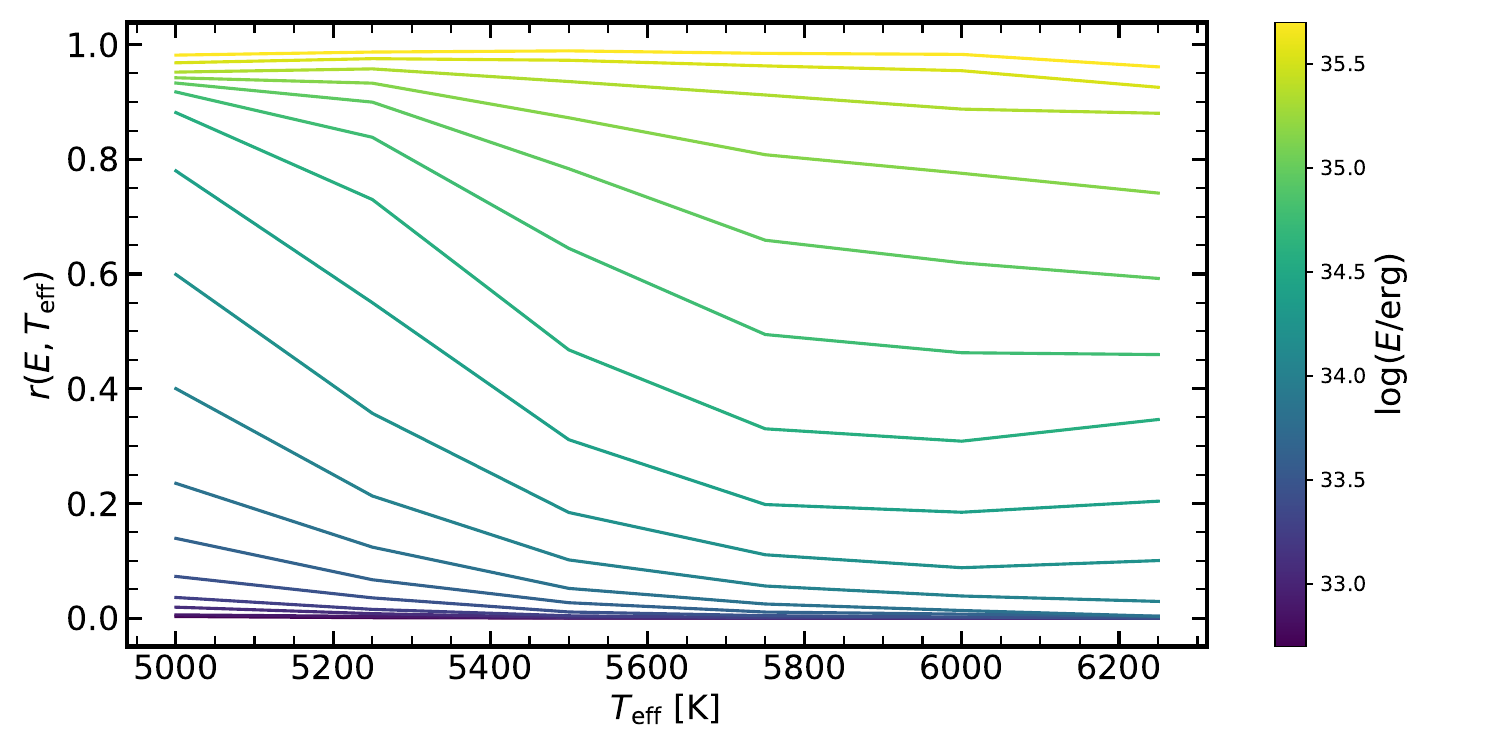}
    \caption{ {\bf Detectability of flares as a function of flare energy and stellar effective temperature.}
    $r (E, \Teff)$ denotes the fraction of detected to injected flares, plotted as a function of stellar effective temperature $\Teff$. Color indicates the flare energy $E$.}
    \label{fig:detectability_Energies_Teff}
\end{figure*}

\begin{table*}[h!]
  \caption{\textbf{Fast rotators excluded from our sample.} Listed are the catalog identifiers (KIC) and  updated ACF periods. These stars were excluded because the period is $<20$ days.}
  \begin{tabular}{cc|cc}
\hline\hline
KIC & $P_{\rm rot, \, ACF}$ (d) & KIC & $P_{\rm rot, \, ACF}$ (d) \\
\hline
2140782 & 0.99 & 7383015 & 16.43 \\
3217852 & 16.96 & 7772296 & 1.19 \\
3426267 & 3.78 & 7840358 & 0.68 \\
3728906 & 3.64 & 7979297 & 12.14 \\
3839928 & 0.95 & 8047700 & 1.08 \\
3971507 & 2.39 & 8142087 & 10.38 \\
4354963 & 6.61 & 9099624 & 1.27 \\
5374537 & 0.51 & 9269566 & 0.95 \\
5534792 & 3.78 & 9290949 & 0.81 \\
5792328 & 12.15 & 9408484 & 0.94 \\
6034834 & 8.63 & 9468023 & 0.90 \\
6041507 & 2.29 & 9515113 & 7.01 \\
6118085 & 1.88 & 9540688 & 0.68 \\
6119243 & 1.08 &  10491884 & 4.64   \\
6119994 & 10.90 & 10534360 & 2.70  \\
6347656 & 14.62 & 10617450 & 2.56  \\
6633602 & 3.38  & 10878191 & 10.59 \\
6790279 & 11.22 & 10972300 & 12.53  \\
6963577 & 14.90 & 11401109 & 0.40  \\
7106500 & 4.86  & 11600387 & 10.94 \\
7109307 & 11.20 & 12303426 & 0.47  \\
7175125 & 13.75 & 12366647 & 19.27 \\
7184946 & 3.64  & \\
7191877 & 3.15 &  &  \\
\hline
\end{tabular}

  \label{Table_3:fast_rot_tab}
\end{table*}

\begin{figure}[ht]
  \centering
\includegraphics[width=1.0\textwidth]{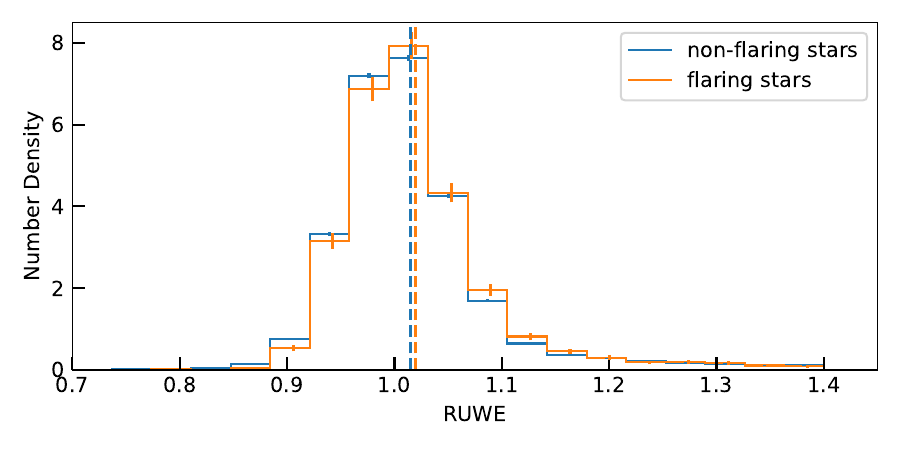}
\caption{\textbf{Distribution of RUWE for flaring and non-flaring stars.} The blue histogram represents non-flaring stars, while the orange histogram represents flaring stars from our stellar sample. The vertical dashed lines show the mean values for both distributions. Vertical error bars indicate $1\sigma$ uncertainties. }
  \label{fig:RUWE}
\end{figure}

\newpage
\noindent
\paragraph*{Supplementary Text}
\paragraph*{Comparison to previous studies.} 
Our analysis confirms previous studies that found superflares do occur on solar-like stars \cite{Maehara2012,Shibayama2013}. Those studies were not sensitive to superflares on stars rotating slower than 20 days, so only found a small number of them  [\!\cite{Shibayama2013}, see their table~2]. That study defined Sun-like stars as stars rotating slower than 10 days. However, it was later found that this definition included many stars much more active and younger than the Sun \cite{Notsu2019, Okamoto2021} which might not reflect the flare occurrence on the Sun.

Other studies of the Kepler data detected multiple flaring stars with rotation periods greater than 20 days, indicating that superflares with energies $>10^{34}$~erg occur on Sun-like stars with near-solar age and activity level once per 2000--3000 years \cite{Notsu2019, Okamoto2021}. This is inconsistent with our result that superflares on such stars occur roughly once per 100 years. In this section we consider possible reasons for this difference.

Our full sample is approximately 34 times larger than in previous work \cite{Okamoto2021}, and we detected approximately 0.05 superflares per star in the sample, which is about three times higher than previous work \cite{Okamoto2021}. We attribute this to the different flare detection algorithms applied to the data. 
To quantify this effect, we applied the previous algorithm \cite{Okamoto2021}  to all \Nflares superflares detected by our method. Of those, only \NumberOfFLaresDetectedWithOkamotoThreshold  superflares satisfied the much stricter detection criteria used in the previous work \cite{Okamoto2021}. This would result in $0.02$ superflares per star, close to their result.

The sample used by previous work \cite{Okamoto2021}  only included stars with known rotation periods determined using the ACF method \cite{McQuillan2014}; stars with no known rotation period were assumed not to generate superflares. Previous work \cite{Okamoto2021} used gyrochronology \cite{Mamajek2008} to estimate how many stars in the Kepler field have near-solar rotation periods. They found that the total number of stars with rotation periods between 20 and 40 days is a factor of 13 larger than the number of observed stars ($N=1641$) with measured rotation periods in this period range. Applying this correction factor reduced their estimate to 0.0012 superflares per star and therefore the flare frequency.

We adopted a different catalog of rotational periods \cite{Reinhold2023} specifically designed for Sun-like stars, which provides the periods for 17,103 stars in our full sample. 
We also included \NstarinputfinalNonPeriodic stars with unknown rotation periods in our analysis.  Our results indicate that stars with Sun-like variability and unknown rotation periods do generate superflares, so no gyrochronology correction is needed (Table~\ref{tab:table_1}).
We therefore did not apply a gyrochronology correction factor.

These effects all increase the frequency of superflares on solar-type stars we measure.
\\

\noindent
\textbf{Relation between flares and SEP events.}
\label{sec:flare-SEP} 
It is not straightforward to compare the occurrence rates and energy estimates of solar flares and SEP events, especially for extreme events  that have not been directly observed \cite{Cliver2022}. 
We expect an indirect relation between extreme SEP events and superflares for two main reasons. Firstly, because the Sun rotates, lines of the solar magnetic field are `frozen' into the radially expanding solar wind. This stretches the field lines into the form of an Archimedean spiral, which magnetically connects Earth to a region on the solar surface near the Sun's west limb \cite{Parker1958, Owens2013}.
If a solar flare occurs in this region, SEPs accelerated during the eruption can reach Earth along the spiral-shaped magnetic field line, producing a SEP event.
If, however, the flare occurs at any other location, the accelerated SEPs would miss Earth.
Even if every flare accelerated particles to high energies, only about one-third of those on the visible hemisphere, those near the west limb, would produce SEPs at Earth \cite{cliver_Apj_2020}.

Secondly, there are no observed constraints on the physical relation between flare energy and the strength of the resulting  SEP event. A statistical relation between the energy of the parent flare and the peak of the event integrated flux (fluence) of SEP measured at Earth is quite broad \cite{cliver20}, introduces an uncertainty of 200 to 300\% in any conversion between flare flux and SEP fluence. This is because multiple physical processes contribute to particle acceleration during a solar flare, which do not closely correlate with flare energy \cite{desai_LR_16}.  Directly observed SEP events are orders of magnitude less intense than those recorded in the cosmogenic isotope record; extrapolating between them might not be valid. For extreme events, the fluxes of SEPs are so intense that feedback mechanisms could modify the acceleration efficiency. This has been proposed on theoretical grounds \cite{reames17}, but it is unknown how it would effect extreme events. For extreme solar flares, we therefore estimate only upper limits on the SEP fluxes expected to be measured at Earth.
Similarly, we estimate only the lower bound of the flare energy from cosmogenic isotopes alone. \\

Data S1: \textbf{Catalog of 56,450 Sun-like stars used in the analysis}. This catalog includes identifiers for each star from both the Kepler mission and Gaia DR3, along with information on the rotation period \cite{Reinhold2023} and stellar parameters \cite{Berger2020}. Detailed description of each column:
\begin{itemize}\setlength{\itemsep}{0.25pt}
    \item \textbf{KIC ID:} Unique identifier from the Kepler Input Catalog.
    \item \textbf{Gaia DR3 Source ID:} Source ID from Gaia Data Release 3.
    \item \textbf{Prot:} Stellar rotation period in days. For stars with unknown rotation period, this value is set to zero. 
    \item \textbf{Teff:} Effective temperature in Kelvin.
    \item \textbf{Teff err:} Mean $1\sigma$ uncertainty in effective temperature, in Kelvin.
    \item \textbf{R:} Stellar radius in units of the Sun's radius.
    \item \textbf{R err:} Mean $1\sigma$ uncertainty in stellar radius, in units of the Sun's radius.
    \item \textbf{log g:} Surface gravity in logarithmic form (logarithmic base 10 of cm$/$s$^2$).
    \item \textbf{[Fe/H]:} Stellar metallicity, in dex.
\end{itemize}

\vspace{0.5cm}

Data S2: \textbf{Catalog of 2889 identified stellar superflares.} The flares were detected in a sample of stars described in Data S1. For each flare, the following columns are included:
\begin{itemize}
    \item \textbf{KIC ID:} Identifier of the flaring Sun-like star, as listed in Data S1.
    \item \textbf{Kepler Time (BJD - 2454844):} The start time of the flare event in Barycentric Julian Date, adjusted by subtracting 2454844 to simplify time representation.
    \item \textbf{Observational Quarter:} Kepler observational quarter during which the flare was observed, ranging from Q0 to Q17.
    \item \textbf{Flare Energy:} Total energy released during the flare, corrected for long cadence data \cite{suppl}, in ergs.
    \item \textbf{E err:} $1\sigma$ uncertainty of the flare energy in ergs.
\end{itemize}

\end{document}